\documentstyle[preprint,aps]{revtex}
\begin{document}
\def\gsim{\hbox{$\lower1pt\hbox{$>$}\above-1pt\raise1pt\hbox{$\sim$}$}}
\def\lsim{\hbox{$\lower1pt\hbox{$<$}\above-1pt\raise1pt\hbox{$\sim$}$}}
\newcommand{\gapro}
  {\raisebox{-0.25ex} {$\,\stackrel{\scriptscriptstyle>}%
    {\scriptscriptstyle\sim}\,$}}
\newcommand{\lapro}
   {\raisebox{-0.25ex} {$\,\stackrel{\scriptscriptstyle<}%
    {\scriptscriptstyle\sim}\,$}}
\title{Polarons and bipolarons in strongly interacting
electron--phonon systems}
\author{{\sc G. Wellein, H. R\"oder and H.~Fehske} \\
 Physikalisches Institut, Universit\"at Bayreuth,
 D--95440 Bayreuth, Germany\\}
\date{Bayreuth, December 19, 1995}
\maketitle
\def\cH{{\cal{H}}}
\def\om{\omega}
\def\cid{c_{i\sigma}^{\dagger}}
\def\cjd{c_{j\sigma}^{\dagger}}
\def\ci{c_{i\sigma}^{}}
\def\cj{c_{j\sigma}^{}}
\def\bid{b_i^{\dagger}}
\def\bjd{b_j^{\dagger}}
\def\bi{b_i^{}}
\def\bj{b_j^{}}
\def\niu{n_{i\uparrow}}
\def\nid{n_{i\downarrow}}
\def\ni{n_i^{}}
\def\nj{n_j^{}}
\def\nis{n_{i\sigma}^{}}
\def\ep{\varepsilon_p}
\def\eb{\varepsilon_b}
\def\eps{\tilde{\varepsilon}_p}
\def\Us{\tilde{U}}
\input{epsf}
\begin{abstract}
The Holstein Hubbard and Holstein t--J models are studied for a wide
range of phonon frequencies, electron--electron and electron--phonon
interaction strengths on finite lattices with up to ten sites by means
of direct Lanczos diagonalization. Previously the necessary
truncation of the phononic Hilbert space caused serious limitations
to either very small systems (four or even two sites)
or to weak electron--phonon coupling, in particular in the adiabatic regime.
Using parallel computers we were able to investigate
the transition from `large' to  `small' polarons in detail.
By resolving the low--lying eigenstates of the Hamiltonian
and by calculating the spectral function we can identify
a polaron band in the strong--coupling case,
whose dispersion deviates from the
free--particle dispersion at low and intermediate phonon frequencies.
For two electrons (holes) we establish the existence of
bipolaronic states and discuss the formation of a bipolaron band.
For the 2D Holstein t--J model we demonstrate that
the formation of hole--polarons is favoured by strong Coulomb
correlations. Analyzing the  hole--hole correlation
functions we find that hole binding is enhanced as a dynamical effect of the
electron--phonon interaction.\\[1cm]
\end{abstract}
\parindent0.8cm
PACS number(s):: 71.27.+a, 71.38.+i, 74.25.Kc, 75.10.Lp
\thispagestyle{empty}
\newpage
\section{Introduction}
Following the discovery of high--temperature superconductivity
in the ceramic copper oxides, novel purely electronic pairing
mechanisms due to the strong Coulomb correlations within the $\rm
CuO_2$ planes  have been investigated in detail.
Recently, however, it has become clear that the lattice degrees of
freedom are essential  in understanding the puzzling
normal--state properties of the cuprates~\cite{BEMB92,Ri94,GC94}.
Even if it should turn out that the electron--phonon (EP) interaction
is not the relevant pairing interaction in those materials,
its effects need to be reconsidered for the case of
strong electron--electron interactions and low effective dimensionality as
realized  in the high--$T_c$ superconductors.
In particular, polaronic effects are suggested to play a
non--negligible role in the copper--based materials
$\rm La_{2-x}Sr_xCuO_{4+y}$~\cite{Ra91,AK92,Emi92b,BE93,AM94,SAL95}  and
even more in the isostructural nickel--based  charge--transfer
oxides $\rm La_{2-x}Sr_xNiO_{4+y}$~\cite{BE93,CCC93}.
Experimentally, photo-induced absorption experiments~\cite{MFVH90},
infrared spectroscopy~\cite{Caea94}
as well as infrared reflectivity measurements~\cite{FLKB93}
unambiguously indicate the formation of `self--localized'
polaronic states (small polarons) in the insulating parent
compounds $\rm La_2CuO_{4+y}$ and $\rm Nd_2CuO_{4-y}$ of the
hole-- and electron--doped superconductors $\rm La_{2-x}Sr_xCuO_{4+y}$ and
$\rm Nd_{2-x}Ce_xCuO_{4-y}$, respectively.
Therefore a growing theoretical interest in the study of
strongly correlated EP  models can be found in the recent
literature~\cite{RT92,Muea92,ZS92,Ra93,AKR94,SYZ95,Feea94,LD94,RFS94,ZL94,FRWM95,DGKR95,Ma95}.

Probably the simplest microscopic models
including both the electron and phonon degrees of freedom are the
Holstein Hubbard model
\begin{eqnarray}
\label{hhm}
\cH_{H-H}=
&&-t \sum_{\langle i j \rangle \sigma}
\Big(c_{i\sigma}^\dagger c_{j\sigma}^{} + {\rm H.c.}\Big)+
U \sum_i n_{i\uparrow}^{} n_{i\downarrow}^{}
\nonumber\\
&&\qquad- \sqrt{\ep\hbar\om}  \sum_i \big(\bid + \bi \big)\,n_i^{}
\;+\;\hbar\om \sum_i \big(\bid\bi + \frac{1}{2}\big)
\end{eqnarray}
and the Holstein t--J model
\begin{eqnarray}
\label{htjm}
\cH_{H-t-J}=
&&-t \sum_{\langle i j \rangle \sigma}
\Big(\tilde{c}_{i\sigma}^\dagger
\tilde{c}_{j\sigma}^{} + {\rm H.c.}\Big)+ J \sum_{\langle i j\rangle}
\Big(\vec{S}_i^{}\vec{S}_j^{} - \frac{1}{4}\tilde{n}_i^{}\tilde{n}_j^{}\Big)
\nonumber\\
&&\qquad- \sqrt{\ep\hbar\om}  \sum_i \big(\bid + \bi \big)\,\tilde{h}_i^{}
\;+\;\hbar\om \sum_i \big(\bid\bi + \frac{1}{2}\big)\,,
\end{eqnarray}
where $c^{\left(\dagger \right)}_{i\sigma}$
annihilates (creates) an electron
at Wannier site $i$ with spin projection $\sigma$, $n_i= n_{i\uparrow}+
n_{i\downarrow}$, and $t$ denotes the transfer amplitude between
nearest--neighbour (NN) pairs $\langle i j \rangle$.
$\cH_{H-t-J}$ acts in a projected Hilbert space without double
occupancy, i.e.,  $\tilde{c}^{\left(\dagger \right)}_{i\sigma}=c^{\left(\dagger
\right)}_{i\sigma}(1- n^{ }_{i\bar{\sigma}})$, and
$\vec{S}^{ }_i=\frac{1}{2} \sum_{\sigma\sigma '}\tilde{c}^\dagger_{i\sigma}
\vec{\tau}_{\sigma\sigma '}^{} \tilde{c}^{ }_{i\sigma '}\;$.
The first two terms in (\ref{hhm}) and (\ref{htjm}) represent the
standard Hubbard model and t--J model, respectively,
where $U$ is the on--site Coulomb repulsion and $J$ measures the NN
antiferromagnetic exchange interaction strength.
The third and fourth terms take into account the EP interaction and the
phonon energy in a harmonic approximation.
Here, the on--site electron (hole) occupation number
$n_i$ ($\tilde{h}_i= 1- \tilde{n}_{i}$) is locally coupled
to a dispersionsless optical phonon mode, where $\ep$ is the EP coupling
constant, $\omega$ denotes the bare phonon frequency,
and  $b_i^{(\dagger)}$ are the
phonon annihilation (creation) operators.
In the context of an effective single--band description of
the copper/nickel oxides, the collective
Holstein--coordinates  $q_i=\sqrt{\hbar/2 M \omega} \, (b_i^\dagger +
b_i^{})$ may be thought of as representing an internal optical degree
of freedom of the lattice site $i$, i.e., in this case the dominant source of
EP
coupling is assumed to result from the interaction of dopant--induced
charge carriers with the apical out--of plane or the
bond--parallel in--plane breathing--type displacements of oxygen atoms.

Unfortunately, for strongly coupled EP systems exact
results exist only in a few special cases and
limits~\cite{Loe88,GL91,CPF95,FL95}.
Whereas, in an approximative treatment, the weak--coupling regime
$(\ep/t\ll 1)$ is  well understood and dealt with by perturbation theory,
the standard strong--coupling Migdal--Eliashberg theory~\cite{Mi58,El60}
based on the adiabatic Migdal theorem might break down for strong
enough EP interactions $(\ep/t\gg 1)$ due to the familiar polaronic band
collapse~\cite{AK92}. Note that in the presence of strong Coulomb
correlations, a rather moderate EP can cause a substantial reduction
of the coherent band motion making the particles susceptible to
`self--trapping'~\cite{ZS92,FRWM95}.
The (single) polaron problem has been tackled in the strong--coupling
adiabatic $(\hbar\omega/t\ll 1)$
and antiadiabatic $(\hbar\omega/t\gg 1)$ limits using the
Holstein~\cite{Ho59a} and Lang--Firsov~\cite{LF62} approximations,
respectively. Both approaches yield a narrow polaronic band
with an exponentially reduced half--bandwidth~\cite{AKR94}.
Whether these small polarons (or bipolarons)
can exist as itinerant band states is still a heavily debated
issue~\cite{BEMB92}.  Apart from variational
calculations~\cite{ZFA89,FCP90,TFDB94,Feea94}
little is known for intermediate values of EP
coupling and phonon frequency $\ep\sim\hbar\omega\sim t$ and, in
particular, for the many--polaron problem. In principle,
exact diagonalization (ED)~\cite{RT92,Ma93,AKR94,Ma95} and
(quantum) Monte Carlo~\cite{RL82,RL83,HF82,NS93,NGSF93}
methods including the full quantum nature
of phonons can close this gap. However, by using direct ED techniques
it is necessary to truncate the phononic Hilbert space,
and hence the accessible parameter space is limited
by the size of the matrix one can diagonalize.
Therefore ED studies up to now were limited to either small
values of $\ep$, to the so--called frozen phonon
approximation~\cite{SZF91,SZ92,FRMB93,RFB93}, or to very small
systems~\cite{RT92,Muea92,AKR94,Ma95}. In a previous work~\cite{FRWM95},
the authors have proposed a variational Lanczos diagonalization technique
on the basis of an inhomogeneous modified variational Lang--Firsov
transformation (IMVLF) that allows for the description of
static displacement field, polaron and squeezing effects in terms of the
Holstein t--J and Holstein Hubbard models on fairly large clusters.
Although the adiabatic and antiadiabatic as well as the weak-- and
strong--coupling limiting cases are well reproduced in this approach,
the situation becomes less favourable at intermediate EP couplings and
phonon frequencies and, in particular, in the crossover region from
large--size nearly free polarons (FP) to small--size `quasi--localized'
polarons (i.e., in the vicinity of the so--called `self--trapping'
transition). Obviously, this regime requires a more accurate treatment
of the phonons as quantum mechanical objects.

Encouraged by this situation it is the aim of the present paper to perform a
direct Lanczos diagonalization of the Holstein Hubbard and Holstein
t--J models, preserving the full dynamics of quantum phonons.
In particular, we investigate for the first time
the low--lying excitations (spectral functions) on large enough
lattices, in order to identify the
dispersion relation of the (bi)polaronic quasiparticles.
\section{Computational Procedure}
A general state of the model Hamiltonian $\cH_{H-H}$ [$\cH_{H-t-J}$]
describing $N_{el}=N_\uparrow+N_\downarrow$ electrons
on a finite $D$--dimensional hypercubic lattice
with  $N$ sites can be written as the direct product
\begin{equation}
\label{gsta}
|{\mit\Psi}\rangle= \sum_{l,k} c_l^k \;
|l\rangle_{el} \otimes |k\rangle_{ph}\,,
\end{equation}
where $ l$ and $k$ label the basic states of the
electronic and phononic Hilbert space with dimensions
$D_{el}={\scriptsize N\choose N_\uparrow}
{\scriptsize N\choose N_\downarrow}$
$\left[D_{el}={\scriptsize N\choose N_\uparrow }
{\scriptsize N-N_\uparrow\choose N_\downarrow}\right]$
and $D_{ph}=\infty$, respectively.
Since the bosonic part of the Hilbert space is infinite
dimensional we use a truncation procedure~\cite{II90}
restricting ourselves to phononic states with at most  $M$ phonons:
\begin{equation}
\label{phonsta}
|k\rangle_{ph}=\prod_{i=1}^N
\frac{1}{\sqrt{n_i^k!}}\left(b_i^\dagger\right)^{n_{i}^{k}}\,|0\rangle_{ph}
\end{equation}
with
\begin{equation}
\label{mabschnei}
\sum_{i=1}^N n_i^k \le M\,,
\end{equation}
and
$1\le k \le D_{ph}^{(M)}=(M+N)!/M!N!\,$.
To further reduce the dimension of the Hilbert space, in the case of
$\cH_{H-H}$  we separate out the center of mass motion
by transforming to new phonon
operators $B_i^{(\dagger)}$, which can be taken into account
analytically as  displaced harmonic oscillators.
For the Holstein t--J model it is more effective
to exploit the  point group symmetries of the original basis~(\ref{gsta}).

Then the  resulting Hamiltonian matrix is diagonalized
using a standard Lanczos method. As the convergence of the
Lanczos procedure depends on the (relative) difference
of neighbouring eigenvalues, $|E_{i+1}-E_i|/|E_i|$,
one needs to be very careful in
resolving eigenvalues within the extremely narrow small--polaron band.
To monitor the convergence of our truncation  procedure as a function of $M$
we calculate the weight of the $m$--phonon states in the ground state
$|{\mit \Psi}_0\rangle$ of $\cH$:
\begin{equation}
\label{cmab}
 |c^m|^2=\sum_{l,k}
|c_{l}^{k}|^2\,,\qquad\mbox{where}\;\;m=\sum_{i=1}^N n_i^{k}\,.
\end{equation}
At fixed $M$, the curve $|c^m|^2 (m)$ is bell-shaped
and its maximum corresponds to the most probable number of
phonon quanta in the ground state. To illustrate the
$M$--dependences of the ground--state energy $E_0$ and the
coefficients $|c^m|^2$, we have shown both quantities
for the single--electron Holstein model in Fig.~\ref{F1}.
In the numerical work convergence is achieved if the relative error of
the ground--state energy is less than $10^{-7}$. In addition, we check
that $E_0$ is smaller than the estimate obtained from the
IMVLF--Lanczos treatment of the phonon subsystem~\cite{FRWM95}.

We have written the program in Fortran90 and ran it on a 64--node
CM5. We were able to diagonalize Hamiltonian matrices up to a total
dimension ($D_{tot}$) of about 82 millions.  Since a matrix vector
multiplication for this matrix size takes less than 150 seconds,
the limiting factor of our numerical algorithm is the available
storage.
\section{Numerical results}
\subsection{Holstein Hubbard model}
\subsubsection{One--electron case}
In the first place, we investigate the polaron properties of the
Holstein model with a single electron on finite lattices
with up to ten sites using periodic boundary conditions.
In the light of the literature over at least the last two
decades~\cite{EH76,RL82,Ma95,GL91,CPF95,FL95} we expect a gradual
transition from a (nearly free) large--polaron solution to a
small--polaron--like ground state upon increasing the EP coupling.
Since, in particular in the adiabatic regime,
the formation of a polaronic state is accompanied by
a strong reduction of the coherent electron motion,
this effect should be observable in the
expectation value of the kinetic energy
$E_{p,kin}/t=-\sum_{<ij>\sigma}\langle{\mit\Psi}_0|(c_{i\sigma}^\dagger
  c_{j\sigma}^{}+\mbox{H.c.})|{\mit\Psi}_0\rangle$, where
$|{\mit\Psi}_0\rangle$ is the ground--state wave-function.
We therefore define an effective polaronic transfer amplitude~\cite{FRWM95},
\begin{equation}
t_{p,eff}=E_{p,kin}(\ep,U)/E_{p,kin}(0,U)\,,
\label{teffpo}
\end{equation}
in order to characterize the increase in the quasiparticle mass~\cite{FRWM95}.
Note that $t_{p,eff}$ substantially differs
from the (exponential) polaron band renormalization factor
$(\rho)$ obtained analytically in the non--adiabatic
Lang--Firsov and adiabatic
Holstein cases~\cite{AKR94}.

We illustrate the dependence of this effective hopping amplitude
on the EP interaction strength in Fig.~\ref{F2},
where we have plotted $t_{p,eff}$ as a function of $\ep$ at different phonon
frequencies. First it is important to realize that there are two complementary
(adiabatic and non--adiabatic) regimes for the polaronic motion.
In the non--adiabatic regime, where the lattice fluctuations
are fast and the phonons are able to follow immediately the electronic
motion forming a non--adiabatic Lang--Firsov polaron (NLFP),
one observes a very gradual decrease of $t_{p,eff}$ as
$\ep$ increases.  At the same time the
`phonon distribution function', $|c^m|^2$, gets wider but the maximum
is still located at the zero--phonon state.
In the adiabatic regime, one
notices a crossover from a large--size polaron (LP) in
1D or nearly free polaron (FP) in 2D,
described by a $t_{p,eff}$ that is only weakly reduced from its
noninteracting value, to a less mobile  (small--size)
adiabatic Holstein polaron (AHP) for large $\ep$.
We point out that the nature of `delocalized' polaronic states,
occurring in the weak--coupling region, is different in
1D and 2D~\cite{FRWM95}. In the 1D case, the FP state becomes unstable
at any finite EP coupling.
As expected the transition to the AHP state occurs
if the EP coupling approximately
exceeds half the bare electronic bandwidth and,
in accordance with Monte Carlo results~\cite{RL82,RL83}, is much
sharper in two dimensions~\cite{We94}
(in the remainder of this section we focus on the 1D case).
Nonetheless, all physical quantities are  smooth functions
of $\ep$, in particular there are no ground--state level
crossings, i.e., the transition from LP/FP to AHP is {\it continuous} and
not accompanied by any non--analyticities. While in the
weak--coupling case we have $m_{max}=0$ and the inclusion of
higher phonon states $(m\gapro 5)$ does not improve the ground--state
energy at all, in the adiabatic strong--coupling  case ($\ep=4$,
$\hbar\omega=0.4$), the maximum in $|c^m|^2$ is shifted to
multi--phonon states ($m_{max}\simeq 8$) and we need about 16 phonons
to reach a sufficient accuracy within our truncation procedure.
Note that a similar behaviour can be observed in
the {\it non--adiabatic} regime
$(\hbar\omega>t)$ provided that $\ep\gg\hbar\omega$, e.g.,
for $\hbar\omega=3$ and $\ep=8$ ($\ep=10$) we find $m_{max}\simeq 2$
in 1D (2D). These results confirm previous findings
for the Holstein Hubbard model on very small size clusters (with
two or three sites), where, as $\ep$ increases
in the adiabatic regime, a strong increase
of the average number of phonons, $\langle N_{ph}\rangle$,
contained in the ground state, was observed (cf. Tab.~I in
Ref.~\cite{RT92} and Tab.~I in Ref.~\cite{Muea92}).
In the center of mass system, the phonon expectation value
in the polaronic ground state may be derived from the
phonon distribution function $|c^m|^2$ by
$\langle N_{ph}\rangle=\sum_{m=0}^M|c^m|^2 +\frac{\ep
}{\hbar\omega} \frac{N_{el}^2}{N}$.

To elucidate the difference between the `extended' LP and
`quasi--localized' AHP states in more
detail, we have calculated the electron--phonon density correlation
function
\begin{equation}
C_{el-ph}^{}(|i-j|)=\langle {\mit\Psi}_0
|n_i^{} b_j^\dagger b_j^{}|{\mit\Psi}_0\rangle\,,
\label{celph}
\end{equation}
which measures the correlation between
the electron occupying site $i$ and the density of
phonons on site $j$~\cite{Ma95b}.
Results for $C_{el-ph}(|i-j|)$,
plotted in Fig.~\ref{F3} at $\hbar\omega=0.4$
for all distances $i-j:=\vec{R}_i-\vec{R}_j$,
show that for small $\ep$ the correlation between the electron and the
phonons is pretty weak and exhibits little structure, i.e., the few
phonons contained in the ground state are nearly uniformly distributed over the
whole lattice. In contrast, in the case of large EP coupling ($\ep=3$),
the phonons are strongly correlated with the position of the electron,
thus implying a very small radius of the polaron.
Note, however, that
the translational invariance of the ground state is not broken. Since a
polaron's mass is inversely proportional to its size,
the AHP formed at large $\ep$ is an extremely heavy quasiparticle.
As can be seen from the inset of Fig.~\ref{F3},
the  on--site electron--phonon correlation increases dramatically
around the same value of $\ep$ at which $t_{p,eff}$ becomes depressed
(cf. Fig.~\ref{F2}).
This means, in the adiabatic regime a strong short--range EP
interaction can lower the energy of the system due to a
deformation--potential--like contribution sufficiently
to overcompensate the loss of kinetic energy.
Nonetheless, the `quasi--localized' (self--trapped) polaronic
state has band--like character, i.e.,  the AHP can move itinerantly.

In order to discuss the formation of a small--polaron band one has to
calculate the low--lying excited states. As a first step, in
Fig.~\ref{F4} we classify the lowest eigenvalues of the Holstein model
according to the allowed wave--vectors of the eight--site lattice for
various phonon frequencies at $\ep=3$.
Here the `band dispersion' $E_K-E_0$ is scaled with respect to the
so--called coherent bandwidth ${\mit \Delta}E=\sup_K E_K -\inf_K E_K$.
${\mit \Delta}E$ strongly depends on
both ratios $\ep/\hbar\omega$ and
$\ep/t$, for example, we found ${\mit
\Delta}E(\ep=3,\hbar\omega)=0.0157$, 0.1957, 2.9165, and 4.0 for
$\hbar\omega=0.4$, 0.8, 10.0 and $\infty$, respectively. Of course,
the simple Lang--Firsov formula,
${\mit \Delta}E_{LF}= 4\mbox{D}  \exp[-\ep/\hbar\omega]$,
gives a good estimate of the polaronic bandwidth
only in the non--adiabatic regime:
${\mit \Delta}E_{LF}(\ep=3,\hbar\omega)=0.0022$
($\hbar\omega=0.4$),  0.0941 (0.8), 2.9633 (10.0), 4 ($\infty$).
Besides the strong renormalization of the bandwidth in the
low--frequency strong--coupling regime
it is interesting to note that the deviation of the polaron band  dispersion
from a (rescaled) `cosine--dispersion'
of noninteracting electrons is most pronounced
at {\it intermediate} phonon frequencies
$\hbar\omega\sim t$, i.e., in between the extreme adiabatic (AHP) and
antiadiabatic (NLFP) limits.
This deviation may be due to a residual polaron--phonon interaction,
with the phonons sitting on sites other than the polaron.
To demonstrate that the low--lying
eigenvalues do indeed form a well--separated quasiparticle band
in the adiabatic strong--coupling regime ($\ep=3$, $\hbar\omega=0.4$),
in the inset of Fig.~\ref{F4} we have displayed
the lowest few eigenvalues in dependence on $\ep$.
In the very weak--coupling regime ($\ep=0.5$) the eigenvalues are
barely changed from their $\ep=0$ values, where additional
eigenvalues, separated from the ground--state energy $E_0$ by multiples of
$\hbar\omega$ (e.g., $E_2$, $E_3$, and $E_4$), enter the spectrum.
As $\ep$ increases a band of states separates from the rest of the
spectrum. These states become very close in energy and a narrow
well--separated energy band evolves in the
strong--coupling case ($\ep=3$). Obviously, the gap
to the next higher band of eigenvalues is of the order
of the bare phonon frequency $\hbar\omega$.
Neglecting degeneracies one may tentatively identify those five
states as the states of the small--polaron band on the eight--site
lattice.

Keeping this identification in mind, in Fig.~\ref{F5} we have
plotted  the lowest eigenvalues as a function of the
(1D) $K$--vectors belonging to various system sizes ($N=6$, 8, 10).
One notices that the dispersion $E_K$ is rather size independent, i.e.,
the $E_K$ values obtained for larger systems just fill the gaps.
Undoubtedly, the smooth shape of $E_K$ already provides good reasons
for a quasiparticle band description of the AHP in the
strong--coupling regime. To further
substantiate this quasiparticle interpretation, we also have
calculated the one--particle spectral functions
\begin{equation}
  A_{K}^{}(E) = \sum_n |\langle {\mit\Psi}_n^{(N_{el})}
|c_{K}^{\dagger}|{\mit\Psi}_0^{(N_{el}-1)}\rangle|^2
\,\delta ( E-E_n^{(N_{el})}+E_0^{(N_{el}-1)})
\label{specfun}
\end{equation}
with $N_{el}=1$ for the non--equivalent $K$--values of the
six--site system using a polynomial moment method~\cite{SR94}.
The idea is to see a direct verification of the
coherent band dispersion $E_K$ in terms of $A_K(E)$.
The electronic spectral functions $A_K(E)$ are shown in
the four insets of Fig.~\ref{F5}.
The important point, we would like to emphasize, is that the position
of the first peak in each spectral function $A_K(E)$ exactly coincides
with the corresponding  $E_K$--value and the other peaks are at higher
energies than any of the coherent band--energy values. This means,
our exact results for the low--energy excitation spectrum of a single
electron corroborate the existence of heavily dressed polaronic
quasiparticles, where the electronic and phononic degrees of freedom
are strongly mixed. Of course,
in the very high--energy regime the results for $A_K(E)$ can not be
trusted just due to the errors induced by the necessary truncation of
the phononic Hilbert space.
\subsubsection{Two--electron case}
Next, we wish to discuss the two--electron problem. Here it is of
special interest to understand in detail the conditions under
which the two electrons form a bipolaron. Whether or not a
transition to a bipolaronic state will occur depends sensitively on
the competition between the short--ranged
phonon--mediated, i.e., retarded $(\hbar\omega
<\infty)$, attraction $(\propto\ep)$
and the instantaneous on--site Hubbard repulsion $(\propto U)$.

We start again with a discussion of the
mobility of the particles. Fig.~\ref{F6} (a) shows the strong
(gradual) reduction of the effective {\it polaronic} transfer amplitude
$t_{p,eff}$ as $\ep$ increases in the adiabatic (non--adiabatic)
regime. Now let us mainly focus on the physically
more interesting regime of `small' phonon frequencies, $\hbar\omega=0.4$.
In the case of  vanishing Coulomb interaction $U=0$
any finite EP interaction causes an effective on--site
attraction between the electrons forming a bipolaronic bound state
(remember, e.g., that $U_{eff}=U-2\ep$ follows from the
simple Lang--Firsov approach). This means, in the pure
Holstein model the state with two nearly free (large)
polarons does not exist, at least in one spatial dimension~\cite{RT92}.
In the weak--coupling limit, the two--polaron state can, however,
be stabilized by taking into account the on--site  Coulomb repulsion.
In this case, a crossover from a state of two mobile large polarons to an
extended bipolaronic state occurs. The `transition' will be
shifted to larger EP couplings as
$U$ increases  (see Fig.~\ref{F6} (a)).
For example, at $U=6$ and $\hbar\omega=0.4$ (3.0), we find that the
binding energy of two electrons,
$E_B^2=E_0(2)-2E_0(1)$, becomes negative at about $\ep=1.7$ (2.8).
Further justification for this interpretation can be found from the
behaviour of the effective bipolaronic transfer amplitude~\cite{IF95},
\begin{equation}
t_{b,eff}=E_{b,kin}(\ep,U)/E_{b,kin}(0,U)
\label{tbipoeff}
\end{equation}
with
\begin{equation}
E_{b,kin}(\ep,U)/t=-\sum_{<ij>}\langle{\mit\Psi}_0(\ep,U)
|(c_{i\uparrow}^\dagger c_{i\downarrow}^\dagger
  c_{j\uparrow}^{} c_{j\downarrow}^{}+\mbox{H.c.})
|{\mit\Psi}_0(\ep,U)\rangle\,,
\label{etbipo}
\end{equation}
shown in Fig.~\ref{F6}~(b). $t_{b,eff}$ describes the coherent
hopping of a on--site bipolaron from site $i$ to site $j$.
Contrary to $t_{p,eff}$, at low EP coupling strengths,
the bipolaronic hopping amplitude $t_{b,eff}$ {\it grows}
with increasing $\ep$ showing the increasing importance of the
correlated motion of two electrons (but, quite clearly, we have
$|E_{b,kin}| < |E_{p,kin}|$). At large EP couplings (e.g., for $\ep\gapro 1$
at $U=0$ and $\hbar\omega=0.4$), the on--site
bipolaron becomes more and more localized and accordingly we observe a
drop in  $t_{b,eff}$ which corresponds to the drop in $t_{p,eff}$ in
the case of one electron at the parameter values where the AHP becomes
stable. Hence we will call this quasiparticle an adiabatic Holstein
bipolaron (AHBP).

To better illustrate the effect of pair formation in the 1D Holstein
(Hubbard) model, we present in Fig.~\ref{F7}
the electron--electron density correlation function
\begin{equation}
C_{el-el}^{}(|i-j|)=\langle {\mit\Psi}_0(\ep,U)
|n_i^{} n_j^{}|{\mit\Psi}_0(\ep,U)\rangle-
\langle {\mit\Psi}_0(0,U)|n_i^{} n_j^{}|{\mit\Psi}_0(0,U)\rangle
\label{celel}
\end{equation}
in the adiabatic regime with (b) and without (a) Hubbard repulsion.
In each case we have displayed the results for $C_{el-el}^{}(|i-j|)$
as a function of $\ep$ in comparison to the electron--phonon
correlation function $C_{el-ph}^{}(|i-j|)$
given by~(\ref{celph}). As Fig.~\ref{F7}~(a) shows,
in the limit of vanishing Coulomb interaction
the on--site electron--electron correlation $C_{el-el}(0)$  dominates
the inter--site correlations $C_{el-el}(|i-j|)$ with $|i-j|\ge 1$, in
particular for $\ep\gapro 0.9$, i.e., in the AHBP regime where both electrons
are mainly confined to the same site sharing a common lattice
distortion. Therefore the transition from a mobile large
bipolaron to a `quasi--self--trapped' on--site AHBP is
manifest in a strongly enhanced $C_{el-ph}(0)$ as well (see inset).
Moreover, the transition should be associated with a significant
reduction of the local magnetic moment, $m_{loc}(\ep,U)\propto
\langle{\mit\Psi}_0|(n_{i\uparrow}-n_{i\downarrow})^2|{\mit\Psi}_0\rangle$,
indicating the local pairing of spin up and down electrons.  Indeed we
found $m_{loc}(\ep=1)/m_{loc}(\ep=0.9)|_{U=0,\hbar\omega=0.4}^{}=0.66$.
As can be seen from Fig.~\ref{F7}~(b), a somewhat
different scenario emerges in the presence
of a finite Coulomb interaction. Here, the Hubbard repulsion
prevents the formation of an on--site bipolaronic bound state
in the weak EP coupling regime. On the other hand, as recently pointed
out by Marsiglio~\cite{Ma95}, the retardation effect of the EP
interaction may favour the formation of more extended pairs.
That is, due to the time--delay
the second electron can take the advantage of the
lattice distortion left by the first one still avoiding
the direct Coulomb repulsion.
In fact, increasing the EP interaction, we find that both the
nearest--neighbour electron--electron and electron--phonon
density correlations starts to rise, while the on--site correlations
remain small (cf. Fig.~\ref{F7}~(b)).
Consequently, we may label this state an adiabatic inter--site bipolaron.
We expect that at larger values of $\ep$ the short--range
EP interaction overcomes the Hubbard repulsion and as a result the
two electrons coalesce on a single site forming a `self--trapped'
bipolaron. Unfortunately we are unable to increase
the dimension of the Hilbert space to contain
a large enough number of phonons in the adiabatic
very strong--coupling regime.

As already mentioned for the one--electron case,
the description of the self--trapping phenomenon requires the
inclusion of multi--phonon states. This is clearly displayed in
Fig.~\ref{F8}, where we have shown the weight of the $m$--phonon
state in the ground state for various EP coupling strengths.
One sees immediately that the maximum of $|c^m|^2$ is rapidly
shifted to larger values of $m$ as $\ep$ increases.
Increasing the phonon frequency at fixed $\ep$,
this tendency is reversed  (see inset).
In the extreme antiadiabatic limit ($\hbar\omega\to\infty$)
we have $m_{max}=0$ and the binding disappears for $U>2\ep$.

As in the case of one electron it is interesting to look at the
low--lying excitations of the inter--site bipolaron. Although we do
not have a clear definition as to the momentum of this compound
particle, it turns out that we indeed find a well--separated energy
band if we again classify the lowest
energy eigenvalues with respect to the allowed $K$--states of our
finite system (see Fig.~\ref{F9}). The formation of the (inter--site)
bipolaron band can be attributed to pronounced retardation effects
[cf. the maxima in the nearest--neighbour correlation functions
$C_{el-ph}(1)$ and  $C_{el-el}(1)$ (Fig.~\ref{F7}) as well as the
large bipolaronic hopping amplitude $t_{b,eff}$ (Fig.~6) at $\ep=3$].
Surprisingly the dispersion of this `quasiparticle' band becomes exactly like
that of a free particle (with a strongly renormalized bandwidth)
at $\ep=U/2$, where in the standard Lang--Firsov polaron theory
the effective Coulomb interaction vanishes. As the EP coupling
exceeds $U/2$, a deviation from the cosine--dispersion occurs and we
expect that for $\ep\gg U/2$ an extremely narrow AHBP--band will be
formed.
\subsection{Holstein t--J model}
Now let us turn to the case, where a few dopant--induced charge
carriers (holes) coupled to lattice phonons move in an antiferromagnetic
correlated spin background. In 2D, this situation, frequently described
by the Holstein t--J model~(\ref{htjm})~\cite{DKR90,FRWM95,DGKR95},
is particularly interesting as it represents the basic electronic and
phononic degrees of freedom in the $\rm CuO_2$ planes of the
high--$T_c$ cuprates. As yet, very little is known theoretically about the
interplay between EP coupling and antiferromagnetic exchange interaction
in such systems. Of course, the exact diagonalization technique, as
applied in the preceding section to the Holstein Hubbard model,
provides reliable results for the ground--state properties of the
Holstein t--J model as well. Here, however, one usually works
near half--filling, i.e., the electronic basis is very
large from the outset imposing severe restrictions on the dimension
of the phononic Hilbert space. Therefore we are unable to reach the
extreme strong EP coupling regime especially in the adiabatic limit.
In the following numerical analysis of the Holstein t--J model, the exchange
interaction strength is fixed to $J/t=0.4$ (which seems to be a
realistic value for the high--$T_c$ systems).

First, let us discuss the behaviour of the effective transfer
amplitude, $t_{p,eff}=E_{p,kin}(\ep,J)/ E_{p,kin}(\ep,0)$, shown in
Fig.~\ref{F10}. Increasing the EP coupling at fixed phonon frequency
$\hbar\omega=0.8$, the mobility of the hole is strongly
reduced and an Holstein--type hole--polaron (AHP) is formed at about
$\ep^c\simeq 2.0$. The continuous crossover
from a nearly free hole--polaron (FP) to
the AHP state is similar to that observed in the 2D single--electron
Holstein model, i.e., at $\ep \simeq \ep^c$ a second maximum in the phonon
distribution function $(|c^m|^2)$ evolves,
which, for $\ep \gg \ep^c$, becomes more pronounced and
is shifted to higher phonon states. For example, we get
$m_{max}\simeq 4$ at  $\ep=4$ and $\hbar\omega=0.8$.
The increasing importance of multi--phonon states in obtaining the
`true' ground--state energy at large $\ep$
becomes clearly visible in Fig.~\ref{F10} by comparing the
results for various phonon numbers $M$.
There is, however, an important difference between the one--hole and
one--electron cases which should not be underemphasized: In the
single--hole Holstein t--J model
antiferromagnetic spin correlations and EP interactions
reinforce each other to the effect of {\it lowering} the threshold for
polaronic `self--localization'.
This fact is in agreement with IMVLF--Lanczos
results obtained recently by the authors~\cite{FRWM95}.
As Fig.~\ref{F10} illustrates, the IMVLF--Lanczos technique, which
variationally takes into account inhomogeneous {\it frozen--in}
displacement--field configurations as well as {\it dynamic}
polaron and squeezing phenomena, describes the qualitative features of
the transition from FP to AHP states and gives a reliable
estimate of the renormalization of the effective transfer
matrix element $t_{p,eff}$.
Moreover, the IMVLF--Lanczos method yields an excellent variational upper
bound for the true ground--state energy $E_0$, and therefore it
provides an additional educated check for the minimal number of
phonons one has to take into account within the Hilbert space
truncation technique.

By analogy to Eq.~(\ref{celph}), we have calculated
the corresponding hole--phonon density correlation function,
$C_{ho-ph}(|i-j|)=\langle {\mit\Psi}_0^{}|\tilde{h}_i^{} b^\dagger_j
b^{}_j|{\mit\Psi}_0^{}\rangle$, for the 2D Holstein t--J model.
Figure~\ref{F11} shows $C_{ho-ph}(|i-j|)$ as a function of the
short--range EP interaction strength $\ep$ at various phonon frequencies.
The transition to the AHP state is signaled by a strong increase in
the on--site hole--phonon correlations which are about one
order in magnitude larger than the nearest--neighbour ones.
This indicates that the AHP quasiparticle comprising a
`quasi--localized' hole and the phonon cloud is mainly confined to a
single lattice site. Increasing the phonon frequency the hole--phonon
correlations are smeared out and the crossover to the small
hole--polaron is shifted to larger values of the EP coupling.

Now, let us consider the two--hole case. In Fig.~\ref{F12} we show
the effective polaronic transfer amplitudes $t_{p,eff}(N_h)$
{\it vs} EP coupling strength in the adiabatic ($\hbar\omega=0.1$),
intermediate ($\hbar\omega=0.8$), and non--adiabatic ($\hbar\omega=3.0$)
regimes. In each case we compare the one-- and two--hole results to
get a feel for hole--binding effects.
Remarkably we find that $t_{p,eff}(2)$ is larger than $t_{p,eff}(1)$
for $\ep\lapro 1$ and $\hbar\omega=0.1$,
indicating a {\it dynamical} type of hole binding in the low--frequency
weak--coupling regime where retardation effects become important.
Indeed, the two--hole binding energy, defined as usual by
$E_B^2(J,\ep,\hbar\omega)=E_0(2)+E_0(0)-2E_0(1)$ with respect to the
Heisenberg energy $E_0(0)$, slightly decreases,
i.e., hole binding is enhanced [$E_B^2(0.4,0,0)<0$], as the EP interaction
increases at low EP coupling strengths.
In contrast, at large phonon frequencies, with increasing $\ep$ we find
that $E_B^2$ increases, which seems to be an indication
that retardation no longer plays a role~\cite{Ma95}.
On the other hand, in the adiabatic strong--coupling
regime, where the two holes become `self--trapped' on NN sites forming
a nearly immobile hole--bipolaron, we expect
an even stronger reduction of $t_{p,eff}(2)$ compared with
$t_{p,eff}(1)$ (cf. the IMVLF--Lanczos results
presented in Ref.~\cite{FRWM95}. Here, a rather {\it static} type of hole
binding is realized.

To substantiate this interpretation we have calculated
the hole--hole density correlation function
\begin{equation}
C_{ho-ho}^{}(|i-j|)=\langle {\mit\Psi}_0(\ep,J)
|\tilde{h}_i^{} \tilde{h}_j^{}|{\mit\Psi}_0(\ep,J)\rangle
\label{choho}
\end{equation}
in the 2D Holstein t--J model. Note that
$C_{ho-ho}(|i-j|)$ provides an even more reliable test for the
occurrence of hole binding than the binding energy
$E_B^2$~\cite{BPS89b}. Indeed, when calculating $E_B^2$, we are
comparing states with different quantum numbers, specifically with
different $S$ and $S^z$. In Fig.~\ref{F13} we present results for
the non--equivalent hole--hole pair correlation functions
in the ground state of the Holstein t--J model with two holes.
In the weak--coupling region the hole--density correlation function
becomes maximum at the largest distance of the ten--site lattice,
while in the intermediate EP coupling regime the preference is on NNN pairs.
As expected, increasing further the EP interaction strength,
the maximum in $C_{ho-ho}(|i-j|)$ is shifted to the
shortest possible distance (remember that double occupancy is strictly
forbidden), indicating hole--hole attraction. The behaviour of
$C_{ho-ho}(|i-j|)$ is found to be qualitatively
similar for higher (lower) phonon frequencies (see inset), except that
the crossings of different hole--hole correlation functions occur at
larger (smaller) values of $\ep$. In essence, our results clearly
indicate that hole--bipolarons could be formed in the Holstein t--J
model at large EP coupling.
\section{Conclusions}
To summarize, in this paper we have studied the problem of (hole--)
bi--/polaron formation in the Holstein Hubbard/t--J model by means of
direct Lanczos diagonalization using a truncation method of the
phononic Hilbert space. Compared with previous treatments of the
Holstein (Hubbard) model on very small clusters,
we are able to analyze large enough systems in order to discuss
polaron and bipolaron band formation, which has been a subject of
recent controversy~\cite{BEMB92,SAL95}. Our main results are the following.
\begin{itemize}
\item[(i)]
In the case of  a single electron coupled to Einstein phonons
(Holstein model), we confirm that
the rather `sharp' transition from a `delocalized'
nearly free polaron (FP)  [or a large polaron (LP) in 1D]
to a `quasi--localized'  Holstein polaron (AHP) in the adiabatic regime
and the very smooth transition to a Lang--Firsov--type polaron (NLFP)
in the non--adiabatic regime are both {\it continuous}.
In agreement with recent exact results~\cite{Loe88,CPF95,FL95},
we observe no ground--state level crossings or any
non--analyticities as the EP coupling increases.
We point out that in the one--dimensional
weak--coupling case a large--size polaron is formed at any finite EP
coupling. In the strong--coupling regime, the AHP state is characterized
by pronounced on--site electron--phonon correlations
making the quasiparticle susceptible to `self--trapping'. Most
notably, the formation of an adiabatic Holstein polaron
is accompanied by a shift of the maximum in the
phonon distribution function to higher phonon states,
which seems to be an intrinsic feature of the
`self--trapping' transition. By contrast, the non--adiabatic
NLFP ground state is basically a zero--phonon state.
\item[(ii)]
By calculating the spectral properties of a single electron, we have
found convincing evidence for the formation of a well separated
narrow polaron band in both the adiabatic and non--adiabatic
strong--coupling regimes. In addition to the expected band--narrowing we
also found a deviation from the `cosine'-dispersion
away from the adiabatic and antiadiabatic limits. Although
the `coherent' bandwidth, deduced from our finite--lattice ED data,
becomes extremely small in the adiabatic
strong--coupling case (polaronic band collapse),
we believe that the AHP does not lose its phase coherence and
can move itinerantly.
\item[(iii)]
Investigating the two--particle problem in terms of the 1D Holstein model,
we could clearly identify the transition
from a extended (large) bipolaron to a `quasi--localized' (on--site)
bipolaron (AHBP) as the EP interaction strength increases.
Stabilizing a two--polaron state in the weak EP coupling regime by
taking into account the on-site Coulomb repulsion (Holstein Hubbard
model), we found a transition to an inter-site bipolaron
at about $\ep\simeq U/2$. It is worth emphasizing that
this inter-site bipolaron appears to have a
dispersion that resembles very closely the cosine--dispersion of a
noninteracting particle with a renormalized bandwidth.
If the EP coupling is further enhanced $(\ep\gg U/2,\,\hbar\omega,\,t)$,
a second transition to a `self--trapped' on--site AHBP
will occur~\cite{We94}.
\item[(iv)]
Analyzing the hole--polaron formation in the framework of the
2D Holstein t--J model, we found that the critical EP coupling for the
polaron transition is substantially reduced due to `prelocalization'
of the doped charge carriers in the antiferromagnetic spin background.
Therefore we suggest that polaronic effects are of special importance
in (low--dimensional) strongly correlated narrow--band
systems like the nickelates and high--$T_c$ cuprates.
\item[(v)]
Regarding ground--state properties of the Holstein t--J model in the
two--hole sector, a detailed study of the hole--hole correlation
functions and the two--hole binding energy was carried out, yielding
strong evidence for an enhanced hole attraction and the formation of
hole--bipolarons as a dynamical effect of the EP interaction.
\end{itemize}

Of course, the exact results presented in this paper hold for
the Holstein Hubbard (Holstein t--J)  model with one and two
electrons (holes) on  {\it finite} 1D (2D) systems, i.e.,
we are not prepared to prove any {\it rigorous} statements
about the thermodynamic limit here. However, we believe that our main
conclusions (i)--(v), in particular the existence of
well--separated polaronic and bipolaronic quasiparticle bands
even in the adiabatic strong--coupling regime,
will survive in the infinite system.
\section*{Acknowledgements}
The computations were performed on a CM5 of the GMD (St. Austin). We thank
D. Ihle and E. Salje for interesting and helpful discussions and J.
Stolze for a critical reading of the manuscript.

\bibliographystyle{phys}

\begin{thebibliography}{10}
\bibitem{Ma95b}
The density--displacement correlation function
has been recently discussed by Marsiglio~\cite{Ma95} for a single electron
on a four--site ring (with 40 phonons) and on a eight--site lattice, but
with a fewer total number of phonons (10); so the author was unable
to check finite--size and truncated Hilbert space effects in the
``highly polaronic'' strong--coupling regime ($\hbar\omega\ll t$).
\end{thebibliography}
\newpage
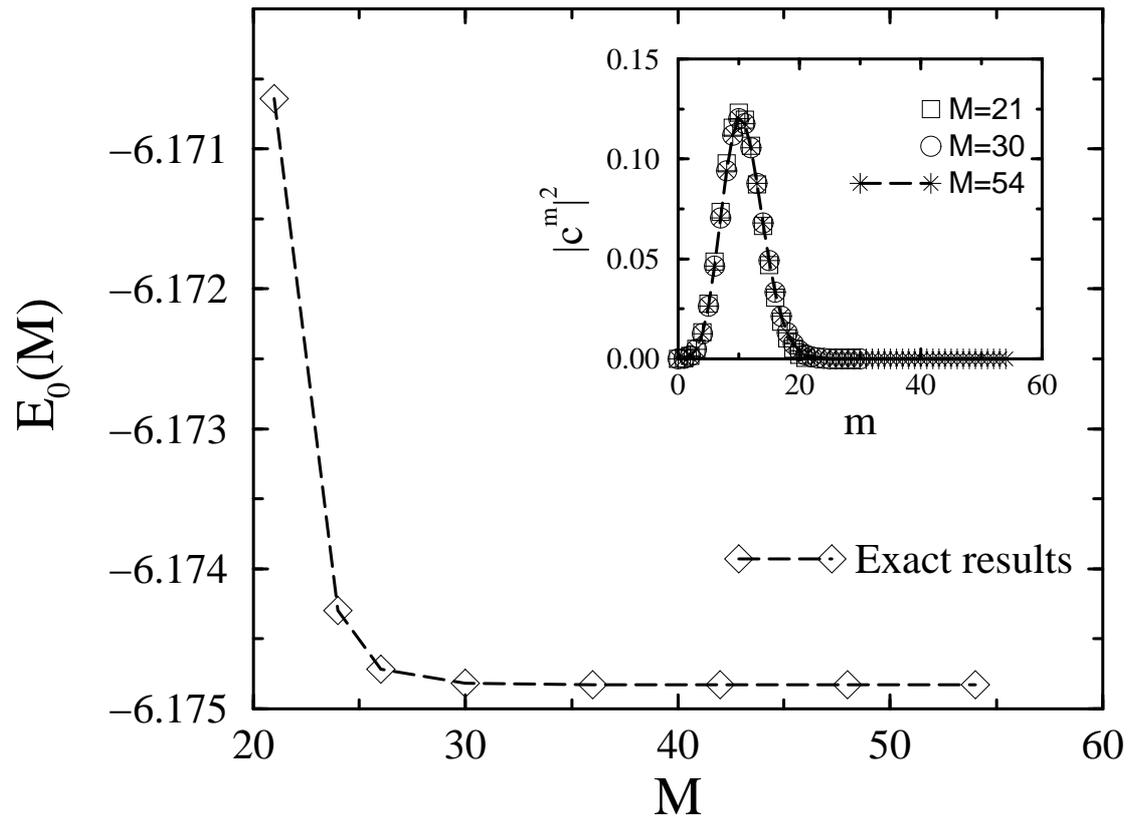
\begin{figure}
\caption{Ground--state energy $E_0$ and weight of the $m$--phonon
states $|c^m|^2$ (inset) as a function of the maximal number of phonons $M$ for
the 1D single--electron Holstein model on a four--site lattice. The
model parameters are: $\ep=6.0$,   $\hbar\omega=0.4$
(all energies are measured in units of $t$).}
\label{F1}
\end{figure}
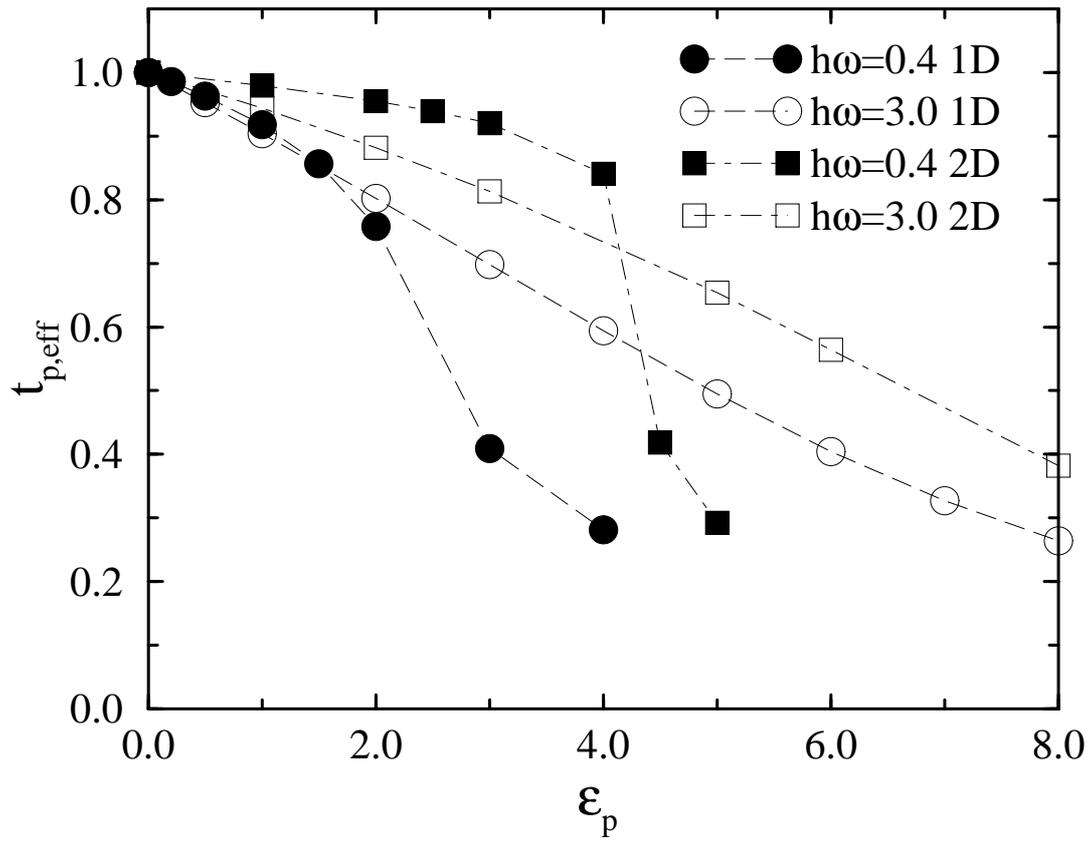
\begin{figure}
\caption{Effective hopping amplitude, $t_{p,eff}$
{\it vs} $\ep$, for a single electron on a ten--site lattice
described by the Holstein model.}
\label{F2}
\end{figure}
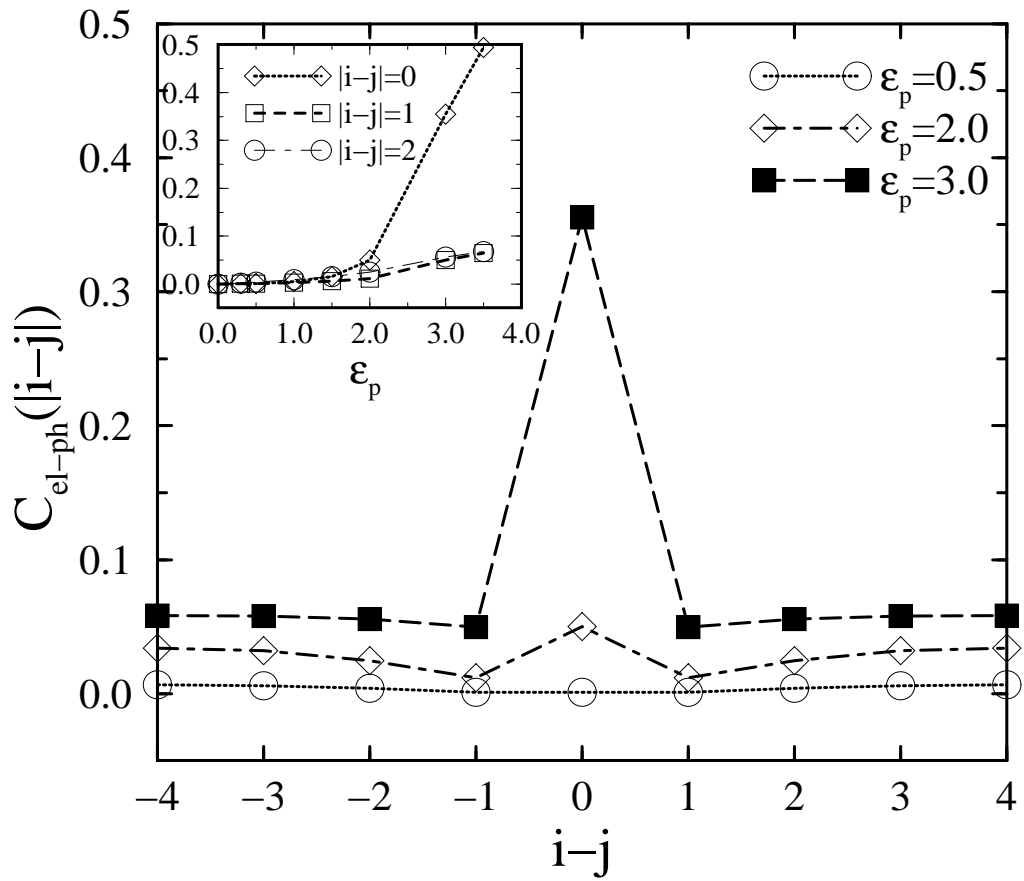
\begin{figure}
\caption{Electron--phonon density correlation function
$C_{el-ph}(|i-j|)$ as a function of the distance $i-j$ and EP coupling
$\ep$ (inset) at $\hbar\omega=0.4$. The results are given for
the single--electron Holstein model on a eight--site chain.}
\label{F3}
\end{figure}
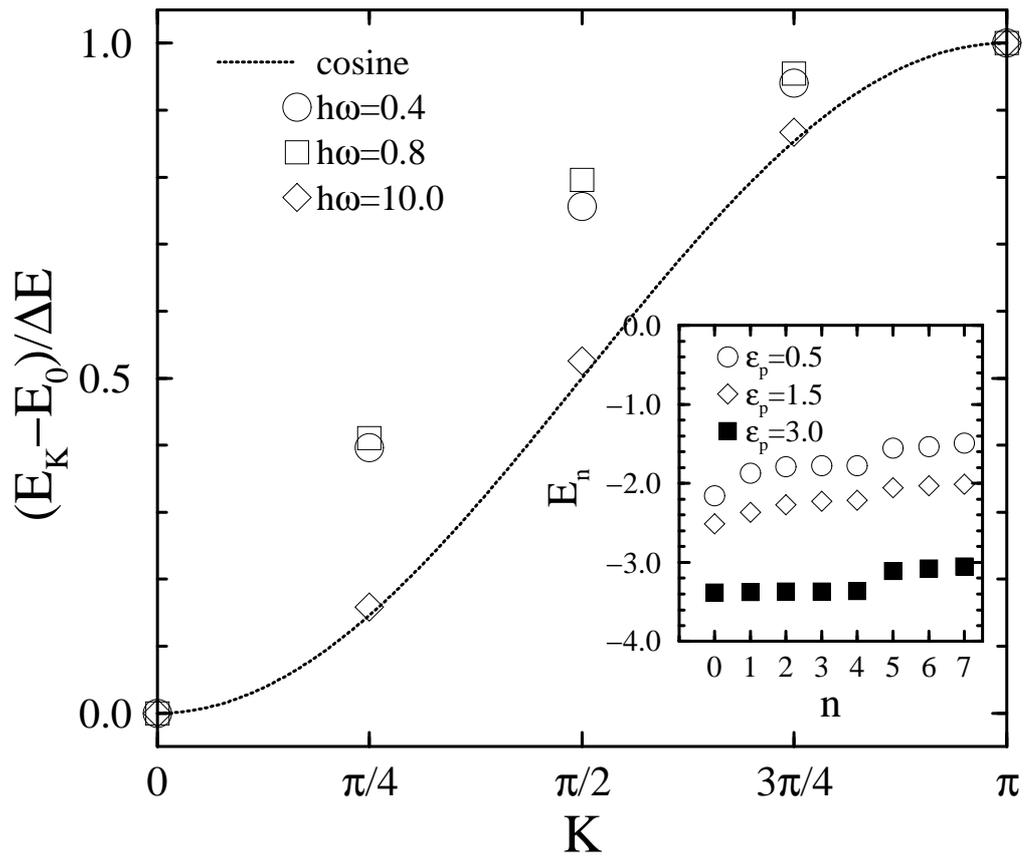
\begin{figure}
\caption{Single--electron band dispersion $E_K-E_0$ at $\ep=3$
in units of the coherent bandwidth ${\mit \Delta}E$.
The inset shows the distribution of the eigenvalues at
$\hbar\omega=0.4$, where the horizontal axis counts the eigenvalues
sorted by magnitude and the vertical axis gives their absolute values
$E_n:=E_n^{(tot)}-N\hbar\omega/2$.
All results are given for the eight--site chain with $M=18$ phonons.}
\label{F4}
\end{figure}
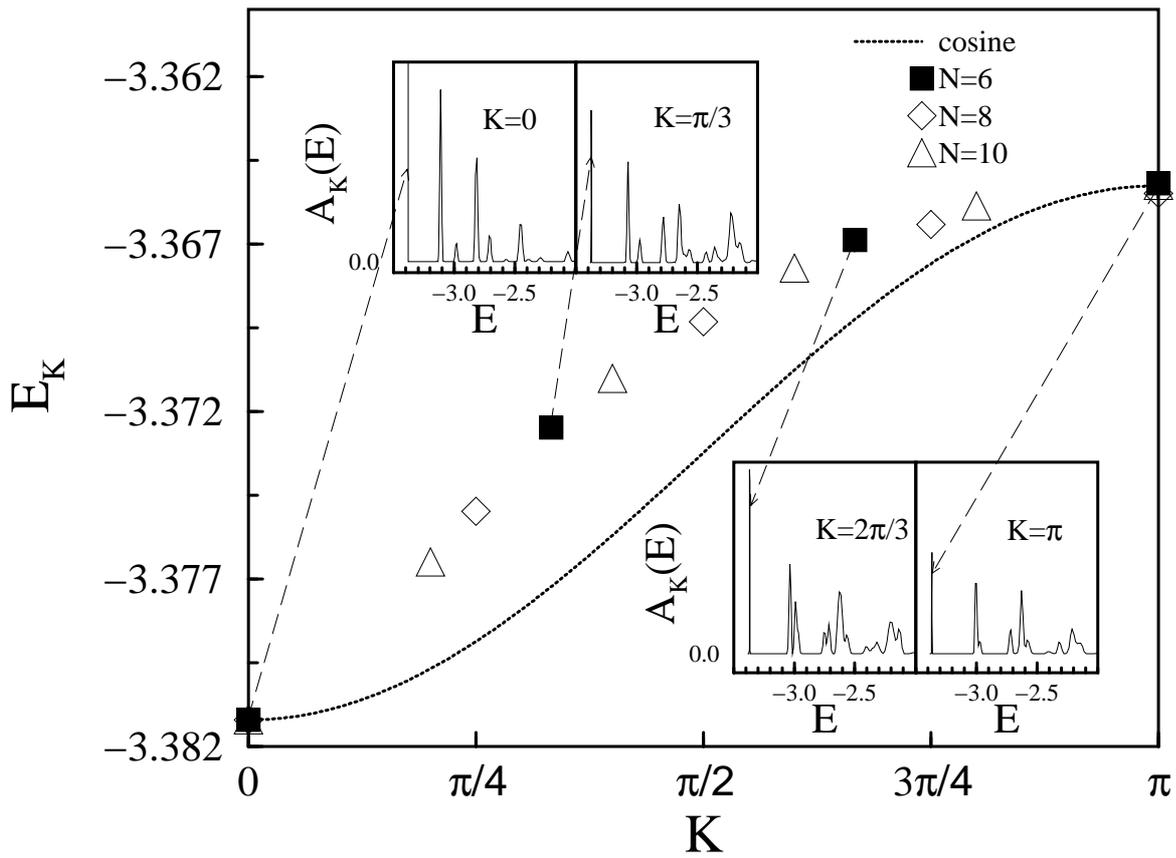
\begin{figure}
\caption{Band dispersion $E(K)$ of a single electron described by
the Holstein model on 1D rings with $N$ sites, where
$\ep=3.0$ and $\hbar\omega=0.4$. The insets
show the low--energy part of the one--particle spectral
function $A_K$(E) taken at the $K$--values indicated by arrows. The
dotted line corresponds to the dispersion of a free particle with a
renormalized bandwidth.}
\label{F5}
\end{figure}
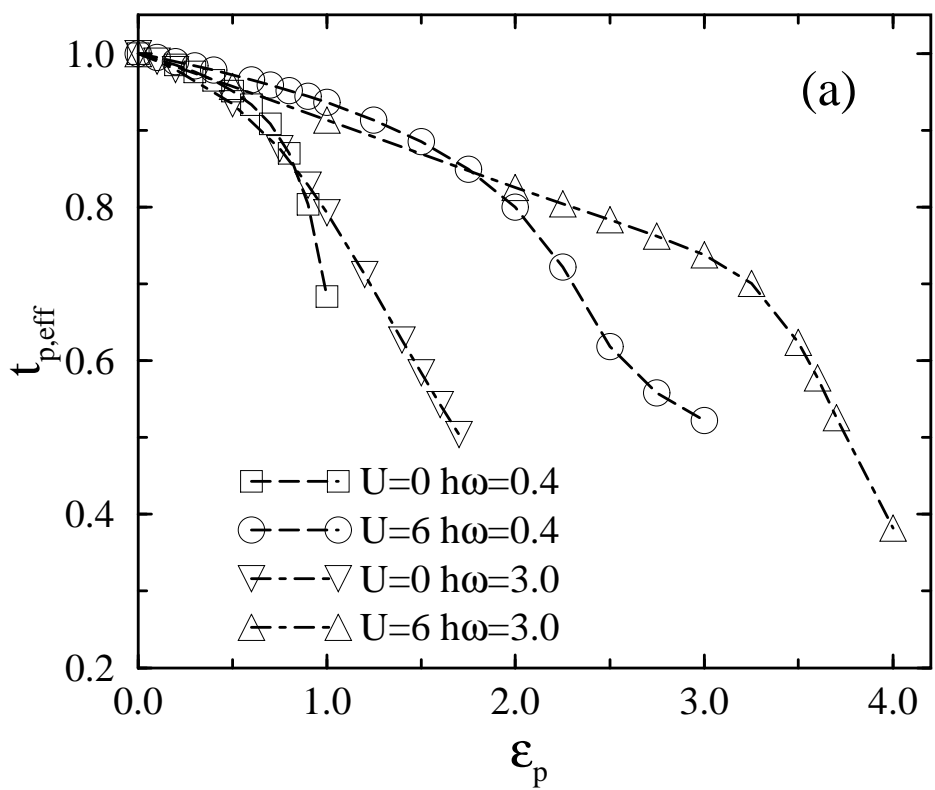
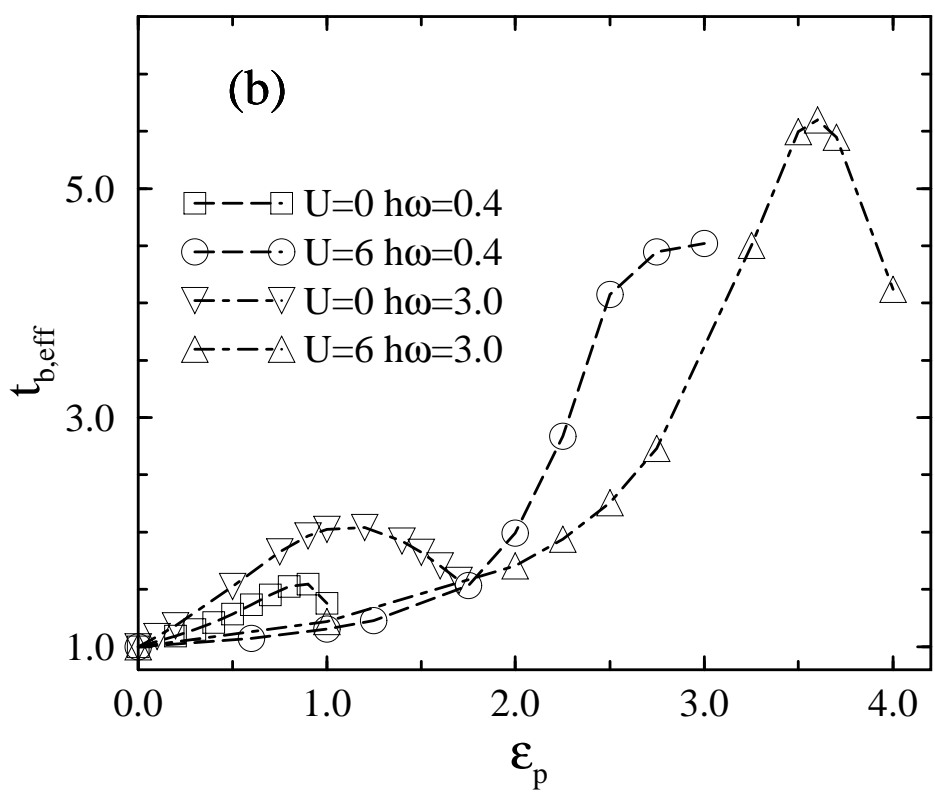
\begin{figure}
\caption{The effective polaron and
bipolaron transfer amplitudes, $t_{p,eff}$ and
$t_{b,eff}$, are shown as a function of the EP coupling strength $\ep$
in (a) and (b), respectively. The results are given
for two electrons on the eight--site chain with $M=21$ phonons using
periodic boundary conditions.}
\label{F6}
\end{figure}
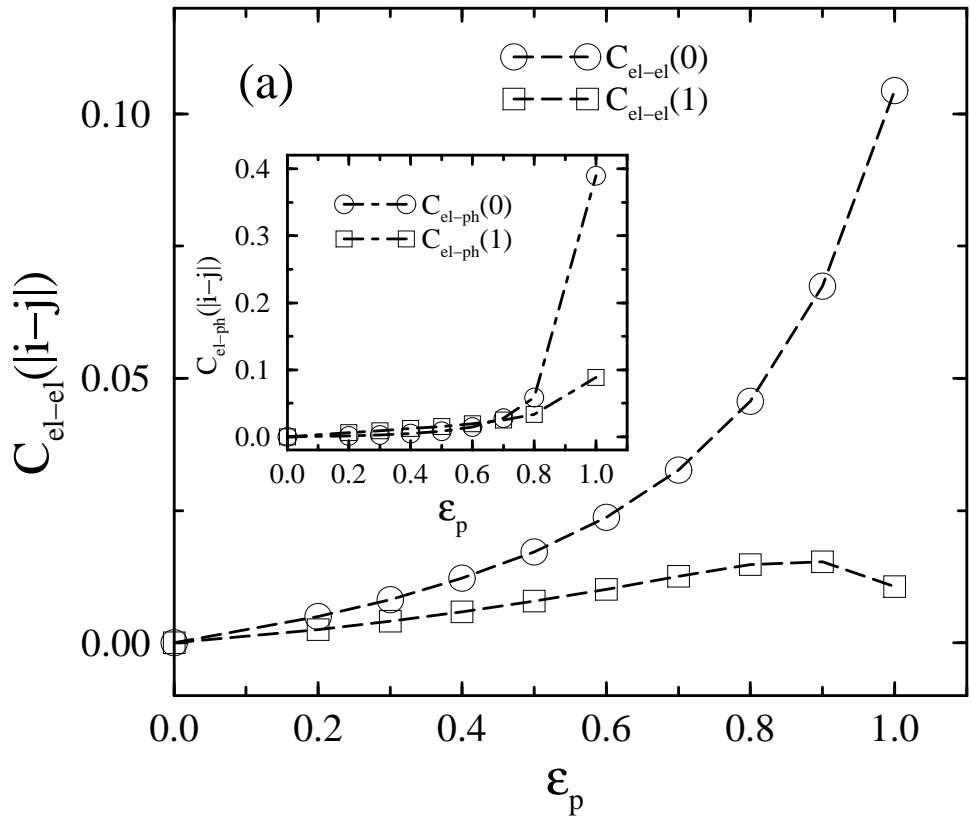
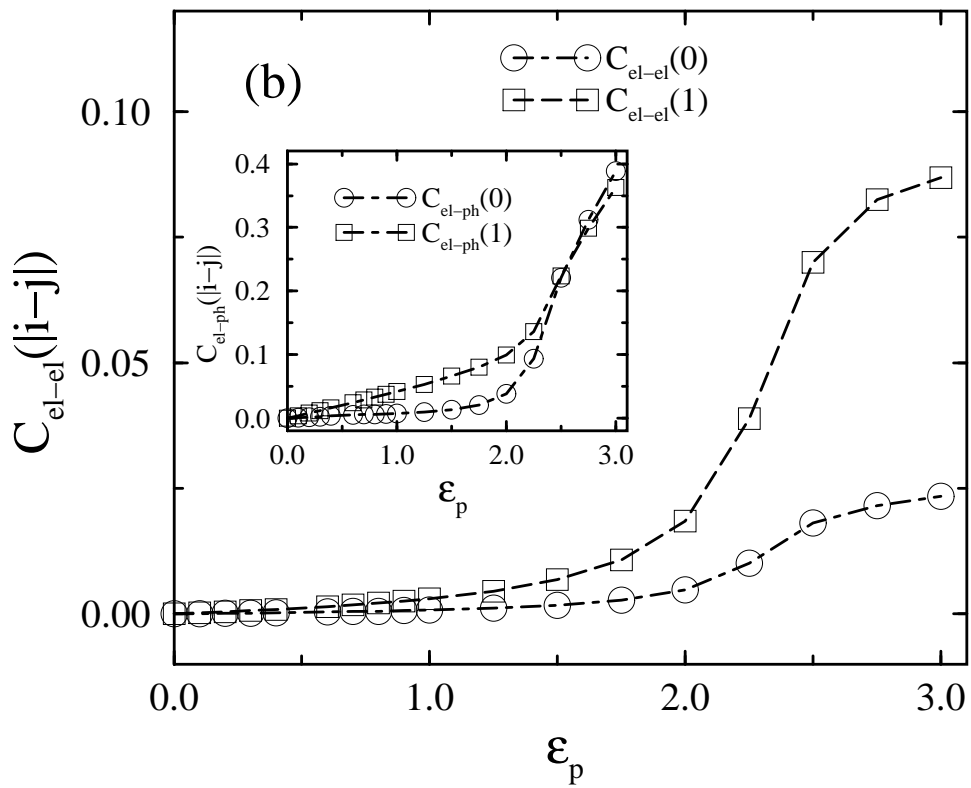
\begin{figure}
\caption{Electron--electron [electron--phonon] correlation function
$C_{el-el}\,$ [$C_{el-ph}$ (inset)] {\it vs} $\ep$
at $\hbar\omega=0.4$ for $U=0$ (a) and $U=6$ (b).}
\label{F7}
\end{figure}
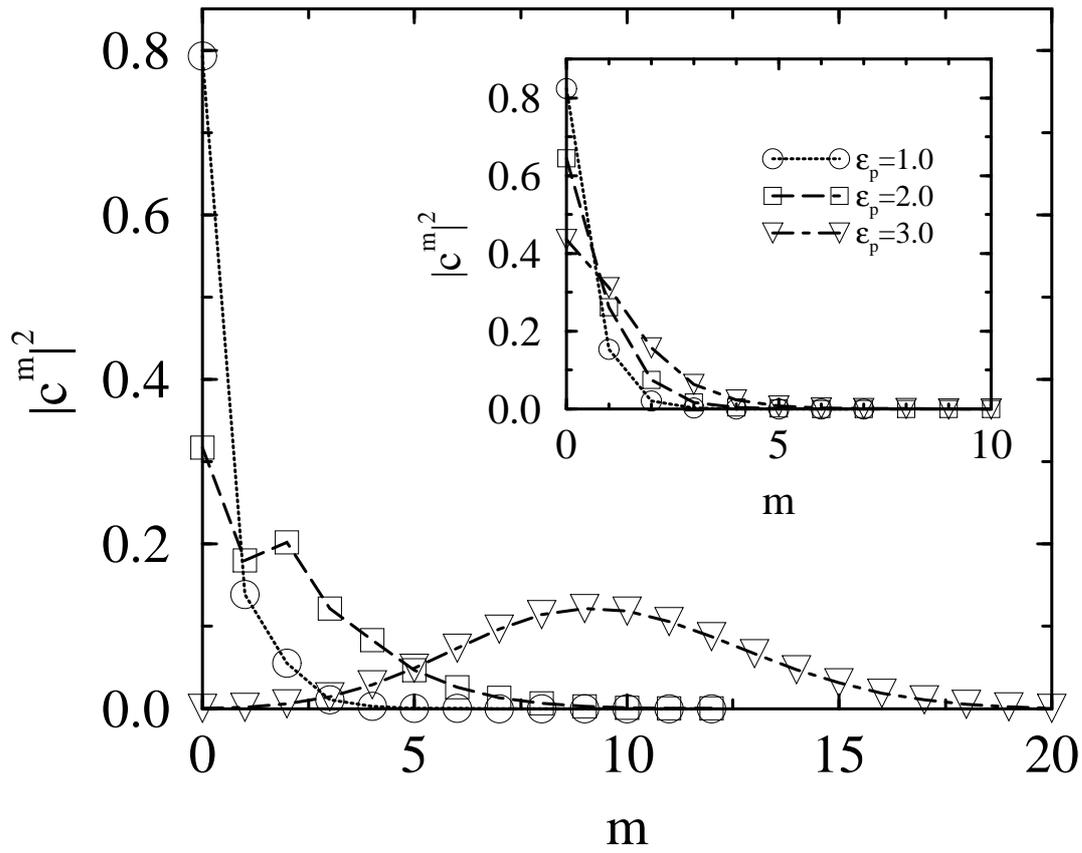
\begin{figure}
\caption{Weight of the $m$--phonon state in the two--electron
ground state of the Holstein
Hubbard model, where $U=6$ and $\hbar\omega=0.4$
[$\hbar\omega=3$ (inset)].}
\label{F8}
\end{figure}
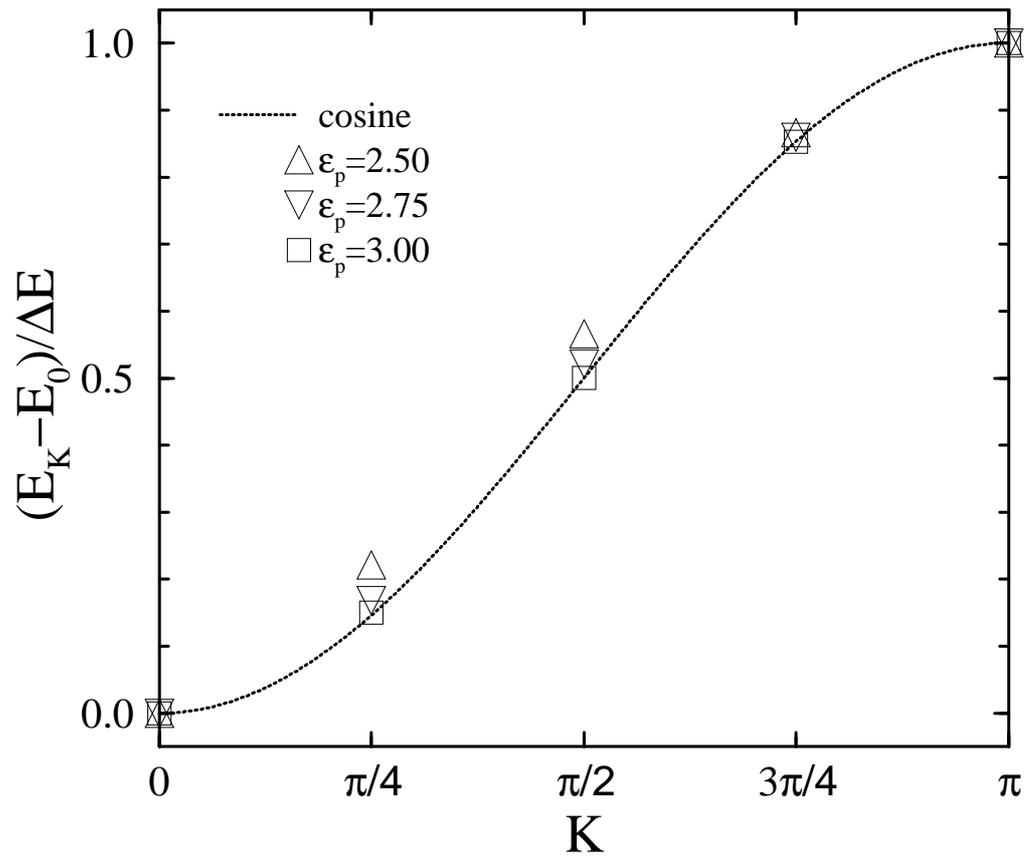
\begin{figure}
\caption{The lowest eigenvalues $E_K$ for the two--electron case on the
eight--site chain, where $\hbar\omega=0.4$ and $U=6$. The energy band
is given in units of the coherent bandwidth ${\mit \Delta} E=0.0481$,
0.0217, and 0.0190 for $\ep=2.5$, 2.75, and 3.0, respectively.}
\label{F9}
\end{figure}
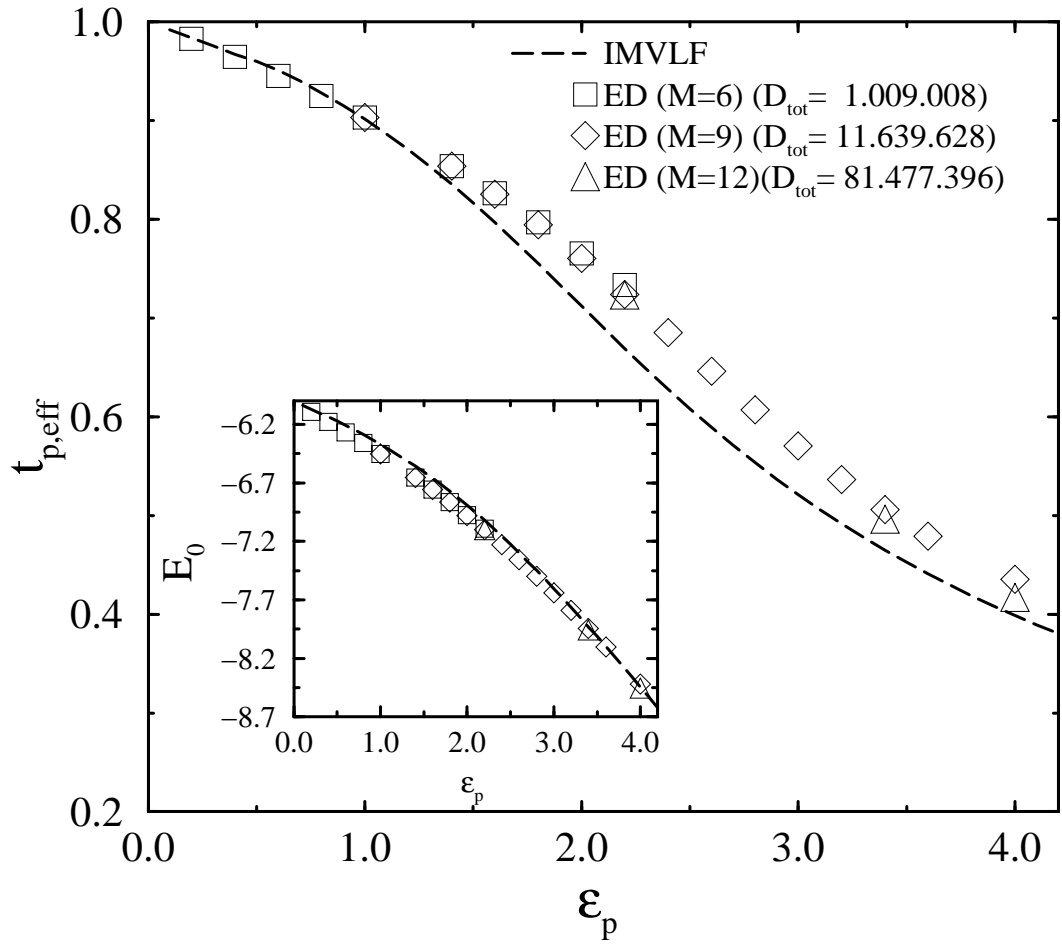
\begin{figure}
\caption{Effective hopping amplitude $t_{p,eff}$ and ground--state
energy $E_0$ (inset) as a function of EP coupling strength $\ep$
for the 2D single--hole Holstein t--J model at $J=0.4$ and
$\hbar\omega=0.8$. Exact results for the ten--site square with different
numbers of phonons $M$ are compared with approximative
IMVLF--Lanczos data~\protect\cite{FRWM95}. For further explanation see text.}
\label{F10}
\end{figure}
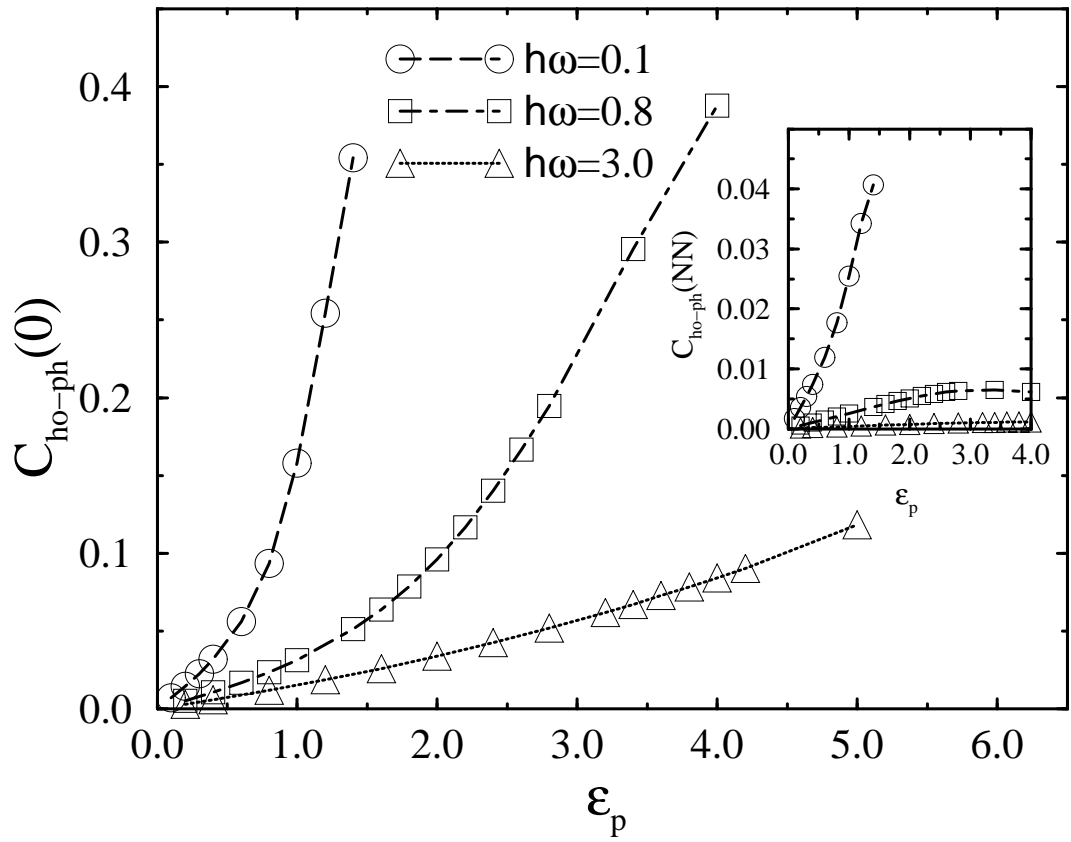
\begin{figure}
\caption{Dependence of the on--site and nearest--neighbour
(inset) hole--phonon correlation function $C_{ho-ph}(|i-j|)$
on the EP coupling $\ep$ and phonon frequency $\hbar\omega$ at
$J=0.4$. Results are given for a single hole
on a ten--site lattice with at most $M=12$ phonons.}
\label{F11}
\end{figure}
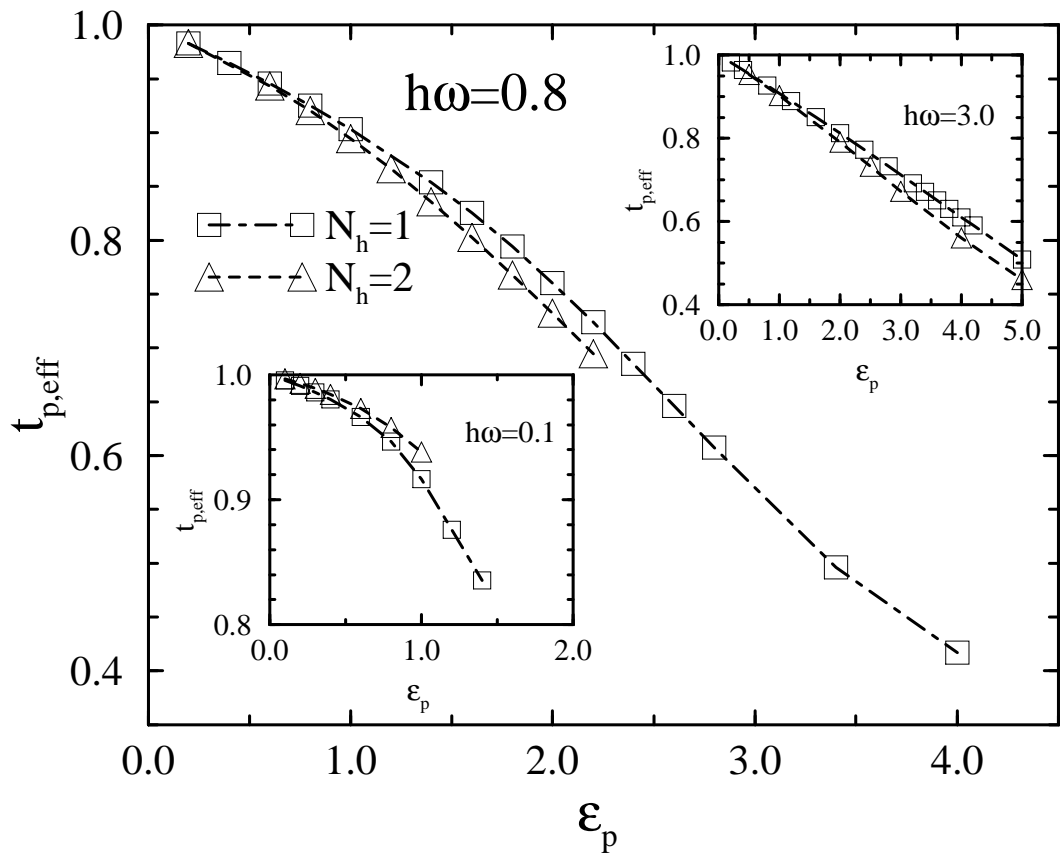
\begin{figure}
\caption{Effective transfer amplitudes, $t_{p,eff}(N_h)$ {\it vs} $\ep$,
are compared for the one-- and two--hole cases
at various phonon frequencies, where $J=0.4$ and $M=12$.}
\label{F12}
\end{figure}
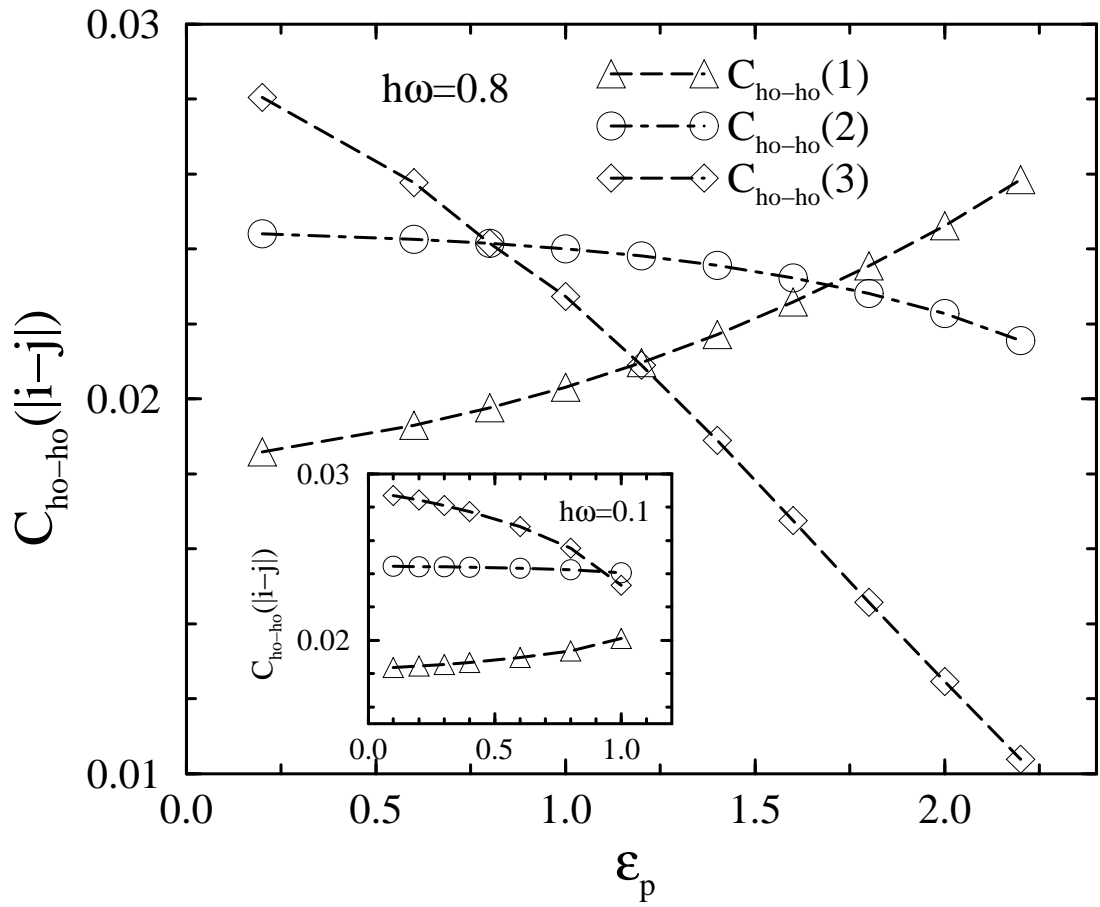
\begin{figure}
\caption{Hole--hole density correlation function $C_{ho-ho}(|i-j|)$
of the 2D Holstein t--J model shown as a function of the EP coupling $\ep$
for $J=0.4$ at $\hbar\omega=0.8$ and $\hbar\omega=0.1$ (inset),
where the numbers 1--3 label the non--equivalent distances
$|\protect\vec{R}_i-\protect\vec{R}_j|/a=1$,
$\protect\sqrt{2}$, and~$\protect\sqrt{5}$
between NN, next NN, and third NN sites on a ten--site lattice with $M=12$
phonons, respectively.}
\label{F13}
\end{figure}
\end{document}